\documentclass[journal]{IEEEtran}
\usepackage{amsmath,amsfonts}
\usepackage{algorithmic}
\usepackage{algorithm}
\usepackage{array}
\usepackage[caption=false,font=normalsize,labelfont=sf,textfont=sf]{subfig}
\usepackage{textcomp}
\usepackage{stfloats}
\usepackage{url}
\usepackage{verbatim}
\usepackage{graphicx}
\usepackage{color}
\usepackage{cite}
\usepackage{booktabs}
\usepackage{multirow}
\usepackage{multicol}
\usepackage{tabularx}
\usepackage{booktabs}  
\usepackage{cleveref}
\usepackage{microtype}  
\usepackage{enumitem}
\usepackage{threeparttable}  
\usepackage{amssymb, makecell, graphicx}

\Crefname{figure}{Fig.}{Figures}
\begin{document}

\title{NEFT: A Unified Framework for Efficient Near-Field CSI Feedback via Knowledge Distillation and Hybrid Design}
 \author{
       Tianqi~Mao,~\IEEEmembership{Member,~IEEE,}
       Haiyang Li,~\IEEEmembership{Student Member,~IEEE,},
       Shufeng Tan,
       Pengyu Wang,\\
       Guangyao Liu,
       Ruiqi Liu,~\IEEEmembership{Senior Member,~IEEE,}
       Leyi Zhang,
       Meng Hua, ~\IEEEmembership{Senior Member,~IEEE,}\\
       Dezhi Zheng, 
       Zhaocheng Wang,~\IEEEmembership{Fellow,~IEEE,}
       and Sheng Chen,~\IEEEmembership{Life Fellow,~IEEE}

\thanks{This work was supported in part by the National Natural Science Foundation of China (NSFC) under Grant numbers 62401054 and 62088101, and in part by Young Elite Scientists Sponsorship Program by CAST under Grant number 2022QNRC001.}
			
\thanks{T. Mao, H. Li and D. Zheng are with State Key Laboratory of Environment Characteristics and Effects for Near-space, Beijing Institute of Technology, Beijing 100081, China. T. Mao is also with Beijing Institute of Technology (Zhuhai), Zhuhai 519088, China (e-mails: maotq@bit.edu.cn; leehaiyang@bit.edu.cn; zhengdezhi@bit.edu.cn).} %

\thanks{S. Tan is with the School of Information and Electronics, Beijing Institute of Technology, Beijing 100081, China (e-mail:3220215117@bit.edu.cn).} %

\thanks{P. Wang and Z. Wang are with the Beijing National Research Center for Information Science and Technology, Department of Electronic Engineering, Tsinghua University, Beijing 100084, China. Z. Wang is also with the Tsinghua Shenzhen International Graduate School, Shenzhen 518055, China (e-mails: wangpengyu@mail.tsinghua.edu.cn; zcwang@tsinghua.edu.cn).} %

\thanks{G. Liu is with the School of Electronic and Information Engineering and the State Key Laboratory of CNS/ATM, Beihang University, Beijing 100191, China (e-mail: liugy@buaa.edu.cn).} %

\thanks{R. Liu is with the Wireless and Computing Research Institute, ZTE Corporation, Beijing 100029, China (e-mail: richie.leo@zte.com.cn).} %

\thanks{L. Zhang is with the Technology Planning Department, ZTE Corporation, Beijing 100029, China (e-mail: leyi.zhang@zte.com.cn).} %

\thanks{M. Hua is with the Department of Electrical and Electronic Engineering, Imperial College London, SW7 2AZ London, U.K. (e-mail: m.hua@imperial.ac.uk).} %

\thanks{S. Chen is with Faculty of Information Science and Technology, Ocean University of China, Qingdao 266100, China (e-mail: sqc@ecs.soton.ac.uk).} %
}


\maketitle

\begin{abstract}
    Extremely large-scale multiple-input multiple-output (XL-MIMO) systems, operating in the near-field region due to their massive antenna arrays, are key enablers of next-generation wireless communications but face significant challenges in channel state information (CSI) feedback. Deep learning has emerged as a powerful tool by learning compact channel features for feedback. However, existing methods struggle to capture the intricate structure of near-field CSI and incur prohibitive computational overhead on practical mobile devices. 
    To overcome these limitations, we propose the near-field efficient feedback Transformer (NEFT) family for accurate near-field CSI feedback with reduced overhead under diverse hardware constraints. 
    NEFT builds on a hierarchical vision Transformer backbone with progressive token reduction and multi-scale feature extraction, enabling compact and effective modeling of near-field channel characteristics.
    Furthermore, NEFT is extended with lightweight variants: NEFT-Compact applies multi-level knowledge distillation (KD) to reduce model complexity while preserving accuracy; NEFT-Hybrid adopts an attention-free CNN encoder to reduce encoder-side computation; and NEFT-Edge combines NEFT-Hybrid with KD to enable deployment on highly resource-constrained edge devices. 
    Extensive simulations show that NEFT achieves a 15--21\,dB improvement in normalized mean-squared error over state-of-the-art methods, NEFT-Compact and NEFT-Edge reduce total FLOPs by 25--36\% with negligible accuracy loss, while NEFT-Hybrid reduces encoder-side complexity by up to 64\%, enabling deployment in highly asymmetric device scenarios. These results establish NEFT as a practical and scalable solution for near-field CSI feedback in XL-MIMO systems.
\end{abstract}


\section{Introduction}\label{S1}

    \IEEEPARstart{A}{s wireless} communication technology is evolving toward the 6th generation, the demand for extremely high data transmission rates and ubiquitous connectivity has driven the continuous development of massive multiple-input multiple-output (MIMO) technology. To meet the stringent performance requirements, the industry is exploring the deployment of extremely large-scale MIMO (XL-MIMO) systems with hundreds or even thousands of antennas\cite{nf_magazine,6G_network,6G_tutorial}. This dramatic expansion in antenna scale brings about an important shift in physical phenomena: as the array aperture significantly increases, the applicable range of traditional far-field plane wave assumption correspondingly shrinks, causing a considerable portion of users to fall within the near-field propagation region\cite{nfris_jstsp,nf_xlmimo_jsac,nf_pi}. Under near-field conditions, the spherical wave characteristics of electromagnetic waves become non-negligible, resulting in channel matrices exhibiting complex nonlinear phase variations and spatially varying amplitude distributions, which contrasts sharply with the relatively simple linear phase relationships in far-field scenarios.
    
    Accurate channel state information (CSI) at the base station (BS) is essential for realizing the gains of massive MIMO systems. In time-division duplexing systems, the BS can exploit uplink-downlink channel reciprocity to obtain downlink CSI from uplink pilots. In contrast, in frequency-division duplexing (FDD) systems, this reciprocity no longer holds because the uplink and downlink operate at different carrier frequencies. As a result, the user equipment (UE) must estimate the downlink CSI locally and feed it back to the BS. When the number of BS antennas becomes very large, this CSI feedback leads to prohibitively high overhead. Therefore, designing efficient CSI feedback mechanisms has become a critical research topic for FDD massive MIMO systems\cite{cf_overview,csinet,crnet,swincfnet,csipppnet,nf_cf_wcl}.
    Most of these works focus on far-field channels, leaving near-field CSI feedback largely unexplored.
    
    While near-field channel modeling\cite{nf_channel_tcom}, estimation\cite{nf_ce_twc,nf_ce_twc2}, and beamforming techniques\cite{nf_bf_cl,nf_bf_wcl} have received considerable attention, little work has investigated CSI feedback mechanisms tailored for near-field channels\cite{nf_cf_wcl}. Due to the complex nonlinear phase and amplitude variations caused by spherical wave propagation, near-field CSI matrices exhibit structural characteristics differing markedly from those in far-field scenarios. As a result, traditional compression methods fail to accurately reconstruct critical channel information under limited feedback overhead. This urgently necessitates the development of feedback solutions specifically tailored for near-field environments.

    \IEEEpubidadjcol
    
\subsection{Related Work}\label{S1.1}

        Traditional CSI feedback methods primarily relied on codebook-based techniques\cite{code_csi} and compressed sensing (CS) approaches\cite{cs_csi,cs_csi2}. Codebook-based methods, deployed in 5G systems through Type I and Type II codebooks, employ pre-designed codebooks shared between the transmitter and receiver. Upon estimating the downlink CSI at the UE, the quantized index of the optimal precoding matrix is computed based on the CSI and the codebook, and then fed back to the transmitter. However, the performance of codebook-based methods is fundamentally limited by the codebook size, which becomes prohibitively large in XL-MIMO systems due to the dramatically increased number of antennas and the complex spatial characteristics of near-field channels. CS-based approaches compressed CSI through linear projections and reconstructed the original channel information by exploiting the inherent sparsity characteristics arising from limited local scatterers in the propagation environment. In near-field environments, however, rich scattering breaks the sparsity assumption, and the iterative reconstruction algorithms incur substantial computational overhead.

        To overcome these limitations, deep learning-based solutions have emerged as promising alternatives for CSI feedback. CsiNet\cite{csinet} and its variants\cite{csinet1,csinet2,csinet3,csinet4,csinet5,csinet6} achieved significant success in far-field environments by learning data-driven representations that outperform CS-based methods across various compression ratios, maintaining effective beamforming gains even at extremely low compression ratios where CS methods failed. 
        However, convolutional neural network (CNN)-based encoders with fixed receptive fields struggle to capture the spatial variations of near-field CSI, and CNN-based decoders have limited capability to model long-range dependencies. 
        ExtendNLNet\cite{nf_cf_wcl} introduced non-local blocks\cite{nonlocal} to capture broader spatial features, but its convolutional backbone still struggled to fully exploiting long-range dependencies and benefit from spatial downsampling, resulting in high computational cost in the fully connected layers.
        
        Recognizing these limitations and motivated by the success of Transformers in computer vision tasks\cite{vit_survey,vit_survey2,vit_survey3,vit_survey4,vit_survey5}, recent studies have explored Transformer architectures for CSI feedback. Building on the Transformer framework\cite{transformer}, TransNet\cite{transnet} adopted a two-layer structure that markedly boosted feedback performance, demonstrating the potential of attention mechanisms for CSI feedback. SwinCFNet\cite{swincfnet} exploited the Swin Transformer\cite{liu_swintf}, achieving further performance gains through window-based multi-head self-attention (W-MSA) and the stacking of multiple attention modules. Nevertheless, due to its direct migration of computer vision architectures while neglecting CSI structural characteristics, the computational cost was further increased compared to TransNet. Therefore, achieving a balance between reconstruction accuracy and computational complexity has became a critical issue in applying Transformers to CSI feedback.
        
        To address the computational constraint, model compression methods, such as pruning, quantization and binarization, have been explored\cite{csi_compress}. With the proliferation of large-scale models, knowledge distillation (KD)\cite{hinton_kd} has emerged as a promising compression approach. KD involves constructing a complex teacher model and a lightweight student model, where the student model is trained to mimic the teacher's output distribution. Preliminary explorations of KD in CSI feedback include methods that introduce temperature parameters to enhance distillation efficiency\cite{csi_kd_conf} and approaches that apply distillation exclusively to encoder components\cite{crnet}. But these methods directly adapt classification frameworks without exploiting the unique structural properties of encoder-decoder architectures, resulting in suboptimal distillation efficiency. Moreover, although self-attention mechanisms have been integrated into CSI feedback, multi-level distillation techniques that jointly leverage attention and codeword information remain unexplored.
        
        Beyond model compression, redesigning autoencoder architectures offers an effective alternative. In next-generation mobile communication systems with Internet of Everything (IoE) deployment, significant hardware asymmetry exists between UEs and BSs. Terminal devices range from intelligent terminals to resource-constrained Internet of Things (IoT) devices, while BSs possess abundant computational resources, necessitating lightweight encoder designs\cite{onesided}. To address this asymmetry, CSI-PPPNet\cite{csipppnet} employed linear mapping at the encoder and combined iterative algorithms with neural networks for CSI reconstruction at the decoder. But its performance is limited by far-field assumption and underperformed compared to baseline methods. Also such one-sided architectures fail account for near-field spatial non-stationarity and spherical wavefront characteristics, making it difficult to balance encoding simplicity with reconstruction accuracy for near-field CSI matrices. Therefore, developing lightweight yet accurate near-field CSI feedback methods remains a critical research challenge.

        The aforementioned DL-based CSI feedback frameworks, such as SwinCFNet\cite{swincfnet}, CSI-PPPNet\cite{csipppnet}, ExtendNLNet\cite{nf_cf_wcl} and KD-based CRNet\cite{crnet}, are mostly developed under the ideal feedback-link assumption, where the compressed codeword is returned without quantization or channel impairments.
        Beyond this assumption, one line of work focuses on the finite-bit encoder–channel interface by introducing quantization strategies to reduce feedback overhead\cite{csinet1,csi_quan1,csi_quan2,csi_quan3}, with more recent studies incorporating differentiable quantization and rate–distortion-oriented objectives into end-to-end optimization\cite{quantization_adaptor,quantization_design,csi_inr}.
        Another line of research addresses the impact of channel noise on the transmitted codeword\cite{csinet1,csi_noise1,csi_noise2}. For instance, the method\cite{csi_noise1} employs a noise extraction module at the BS and adopts joint training to improve the robustness of CSI reconstruction.
        Inspired by deep joint source–channel coding, ADJSCC-CSINet\cite{adjscc_csi} integrates source compression and noisy-channel adaptation by coupling non-linear transforms with noise-aware processing for CSI feedback.
        
        While these works enhance practical robustness through quantization and channel-aware designs, many studies still concentrate on improving the intrinsic representation capacity of the encoder–decoder backbone, as stronger feature extraction and compression capability can naturally accommodate additional modules such as quantization or denoising. Enhancing the structural expressiveness of the core network remains a key unmet requirement.

 \subsection{Motivation and Contribution}\label{S1.2}

        Motivated by the above observations and the lack of near-field–oriented encoder–decoder designs under realistic hardware asymmetry, we propose the near-field efficient feedback Transformer (NEFT) family, a unified framework for near-field CSI feedback. 
        Our approach combines novel architectural designs with advanced KD strategies to enable practical deployment across diverse hardware configurations.
        The main contributions of this paper are highlighted as follows.

\begin{enumerate}
            \item We develop NEFT, a hierarchical vision Transformer (ViT)-based model with multi-stage downsampling and upsampling to balance global attention computation and memory efficiency. The model is tailored to capture near-field spherical wave characteristics, enabling precise CSI feedback on high-performance intelligent terminals.
            \item We propose a multi-level alignment KD strategy to derive NEFT-Compact from NEFT. By jointly aligning attention maps, codewords and reconstruction outputs, the framework preserves near-original performance on resource-constrained devices while significantly reducing model complexity.
            \item We present NEFT-Hybrid, integrating a lightweight CNN encoder with the NEFT decoder for encoder-constrained scenarios. We further develop NEFT-Edge, applying KD to enable highly resource-constrained IoT deployment. This cascaded design supports the full hardware spectrum, from high-performance servers to edge devices.
            \item We conduct extensive evaluations to demonstrate the superior performance of the NEFT family across diverse hardware platforms. Results show that NEFT achieves a 15--21\,dB improvement in normalized mean squared error (NMSE), while NEFT-Edge surpasses existing methods with higher computational efficiency.
\end{enumerate}

\begin{table}[h]
\vspace*{-3mm}
\caption{Qualitative comparison between our NEFT family and representative CSI feedback methods.}
\label{tab:comparison_single} 
\vspace*{-4mm}
\begin{center}
\begin{threeparttable}
    \renewcommand{\arraystretch}{1.05} 
    \setlength{\tabcolsep}{4pt}        
    \begin{tabular}{lcccc}
    \hline
    \textbf{Method} &
    \makecell{\textbf{Near-field}\\\textbf{Support}} &
    \makecell{\textbf{Storage}\\\textbf{Req.}} &
    \makecell{\textbf{Comp.}\\\textbf{(Full)}} &
    \makecell{\textbf{Comp.}\\\textbf{(Enc.)}} \\
    \hline
    CsiNet~\cite{csinet}          & $\times$        & Mod.      & Mod.        & Low\\
    ExtendNLNet~\cite{nf_cf_wcl}  & $\checkmark$& Mod.& High--Huge& Low--Mod.\\
    SwinCFNet~\cite{swincfnet}    & $\checkmark$    & Huge & Huge   & Huge \\
    \hline
    \textbf{NEFT (ours)}          & $\checkmark$    & Mod.--High& High& High\\
    \textbf{NEFT-Compact}         & $\checkmark$    & Mod.& Mod.& Mod.  \\
    \textbf{NEFT-Hybrid}          & $\checkmark$    & Mod.& Mod.        & Low--Mod.\\
    \textbf{NEFT-Edge}            & $\checkmark$    & \textbf{Low}  & \textbf{Low} & \textbf{Low}   \\
    \hline
    \end{tabular}
    \begin{tablenotes}
    \footnotesize
    \item[] $\checkmark$ — fully supported; $\times$ — not supported. Mod. — Moderate; ``Huge'': significantly larger ($>10\times$) in parameters/computing cost.
    \end{tablenotes}
    \end{threeparttable}
\end{center}
\vspace*{-2mm}
\end{table}

    Table~\ref{tab:comparison_single} provides a qualitative comparison between our NEFT family and representative CSI feedback schemes, including near-field adaptability, storage requirements, and both overall and encoder-side computation costs, to highlight NEFT's suitability for diverse hardware constraints. Since SwinCFNet exhibits an order-of-magnitude higher computational and storage complexity than other schemes, it is therefore excluded as a baseline in the experiment section.

\section{System Model and Problem Formulation}\label{S2}

\subsection{Near-Field Channel Model}\label{S2.1}

    We consider a single-cell massive MIMO system operating in FDD mode. The BS is equipped with a uniform linear array (ULA) of \(N_1\) antennas and communicates with multiple UEs, each having \(N_2\) antennas, all located within the near-field coverage area of the BS. The received signal at a given user is modeled as\footnote{We assume that the system adopts orthogonal access. Alternatively, for non-orthogonal access, the precoding vector for a particular user is designed to be orthogonal to other users' channel matrices.}
    \begin{equation}\label{eq1}
      \mathbf{y} = \mathbf{H}\mathbf{v}x + \mathbf{n},
    \end{equation}
    where \(x\) is the transmitted symbol, \(\mathbf{v}\in\mathbb{C}^{N_1\times1}\) is the precoding vector, 
    \(\mathbf{H}\in\mathbb{C}^{N_2\times N_1}\) denotes the channel matrix, 
    and \(\mathbf{n}\in\mathbb{C}^{N_2 \times 1}\) represents the additive noise vector at the receiver.

    The near-field assumption in our system model fundamentally alters the propagation characteristics compared to conventional far-field communications. In far-field scenarios, the large transmitter-receiver separation allows incoming waves to be approximated as plane waves, resulting in linear phase variations across antenna elements. However, when the communication distance falls below the Rayleigh distance\cite{Rayleigh}
    \begin{equation}\label{eq2}
      d_R = \frac{2D^2}{\lambda},
    \end{equation}
    where $D$ is the maximum antenna array dimension and $\lambda$ is the wavelength, the wavefront becomes spherical. In systems with a large number of antennas operating at extremely high frequencies, $d_R$ is reduced significantly, thereby placing most users in the near-field region. This spherical propagation leads to nonlinear phase and amplitude variations across the antenna array.
    To characterize this near-field propagation, we employ a geometric free-space line-of-sight (LoS) model, which is widely adopted in near-field MIMO studies and near-field CSI feedback research\cite{nf_channel_tcom,nf_cf_wcl}. Specifically, the channel coefficient between the $n_1$-th BS antenna and the $n_2$-th UE antenna is given by
    \begin{equation}\label{eq3}
      \mathbf{H}(n_1,n_2) = \frac{1}{r_{n_1,n_2}}\, \exp\!\left(-j\frac{2\pi}{\lambda}\,r_{n_1,n_2}\right),
    \end{equation}
    where $r_{n_1,n_2}$ denotes the physical propagation distance between antenna pairs, which can be expressed as
    \begin{equation}\label{eq4}
       r_{n_2,n_1}\!\! =\!\! \sqrt{\! (r\cos\theta\! -\! d_2\sin\phi)^2\! +\! (r\sin\theta\! +\! d_2\cos\phi\! -\! d_1)^2}.\!
    \end{equation}
    In (\ref{eq4}), $r$ is the distance between the first BS and UE antennas, $\phi$ denotes the relative angle between UE and BS, and $\theta$ denotes the angle of departure (AoD) of the signal, while $d_1 = n_1 d$ and $d_2 = n_2 d$, with $d$ being the antenna spacing.

    This geometric model captures the distance-dependent path loss and phase variations of near-field propagation. The resulting CSI matrices exhibit complex spatial structures with nonlinear phase relationships that differ significantly from conventional far-field patterns, presenting unique challenges for efficient compression and feedback.

\subsection{CSI Feedback in Near-Field}\label{S2.2}

    Given the increased complexity of near-field CSI matrices, efficient feedback mechanisms are critical in FDD systems. Accurate precoding and beamforming at the BS require the UE to first estimate the CSI matrix \(\mathbf{H}\) using pilot signals and then feed back a compressed version. To focus on the feedback mechanism, we assume that the CSI is perfectly acquired via pilot-based estimation.

    To address the complexity of near-field CSI compression, deep learning-based approaches have emerged as a promising solution. These schemes typically employ an end-to-end learning framework, where an encoder $\mathcal{E}(\cdot)$ and a decoder $\mathcal{D}(\cdot)$ are jointly optimized to minimize reconstruction error under compression constraints. In this framework, the CSI matrix is decomposed into its real and imaginary components and concatenated to form a two-channel real-valued matrix \(\mathbf{H}_{\text{in}}\) of size $L = 2N_2N_1$ with all elements normalized to the range \([0,\, 1]\). 
    The encoder compresses $\mathbf{H}_{\text{in}}$ into a $K$-dimensional feedback codeword $\mathbf{s}$
    \begin{equation}\label{eq5}
      \mathbf{s} = \mathcal{E}(\mathbf{H}_{\text{in}}),
    \end{equation}
    which is assumed to be transmitted over a perfect feedback link as in\cite{csipppnet,crnet}, and the decoder at the BS reconstructs an approximation $\hat{\mathbf{H}}$ of the original CSI matrix:
    \begin{equation}\label{eq6}
      \hat{\mathbf{H}} = \mathcal{D}(\mathbf{s}).
    \end{equation}

    The compression rate is defined as the ratio of the compressed codeword size to the original CSI size:
    \begin{equation}\label{eq7}
      \gamma = \frac{K}{L} = \frac{K}{2N_1N_2}.
    \end{equation}
    However, achieving effective compression in near-field scenarios requires addressing unique architectural challenges. Specifically, the design objective is to achieve optimal reconstruction fidelity under stringent feedback constraints while maintaining precoding performance. Near-field scenarios pose unique challenges due to complex spatial patterns with nonlinear phase relationships that differ significantly from those in far-field communications. To address these challenges, encoders must capture both local details and global nonlinear features through large receptive fields while preserving critical positional information. Decoders require robust modeling capabilities to reconstruct fine-grained local patterns and long-range dependencies, often necessitating advanced sequence processing architectures.

\section{NEFT: Hierarchical Vision Transformer}\label{s3}
 
    This section presents NEFT, a hierarchical ViT architecture designed for efficient near-field CSI feedback.

    \begin{figure*}[!t]
      \centering
      \includegraphics[width=\textwidth]{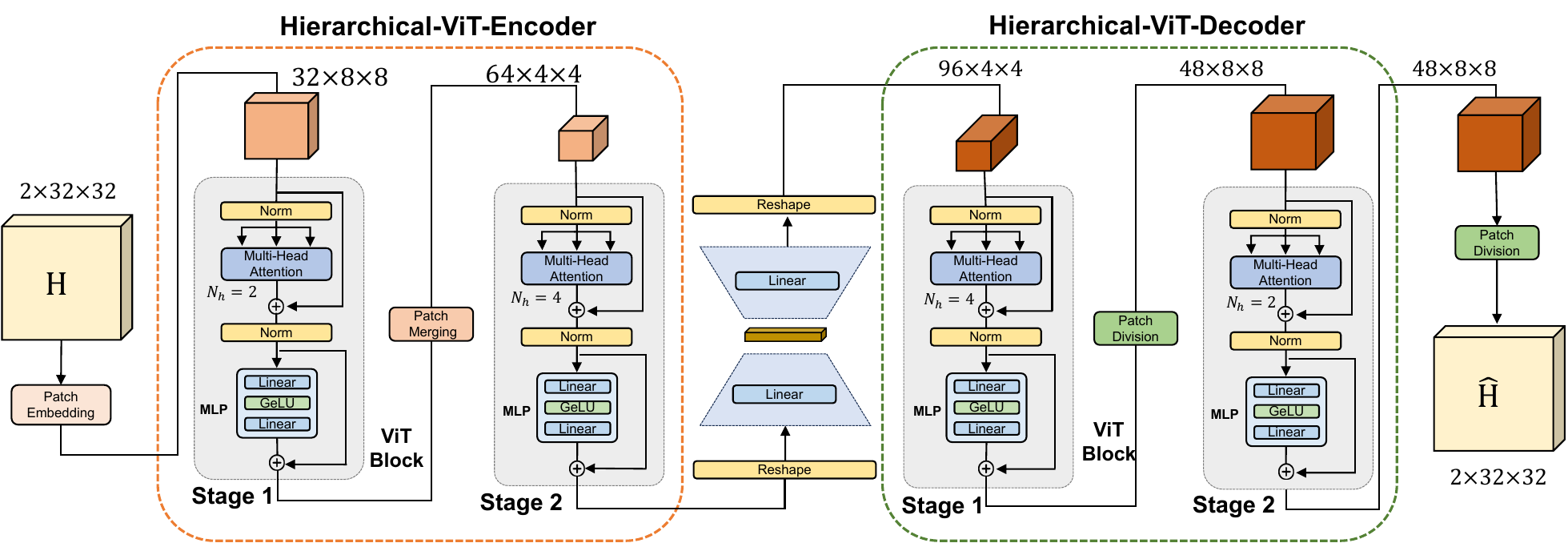}
      \caption{The proposed hierarchical Vision Transformer (HViT) architecture with NEFT. The two-stage encoder-decoder structure employs global attention mechanisms across progressively downsampled feature maps, achieving computational efficiency while maintaining full spatial dependency modeling capabilities for near-field CSI feedback applications. The parameter configuration shown corresponds to the case of $\gamma = 32$.}
      \label{fig:NEFT} 
			\vspace*{-4mm}
    \end{figure*}
    
\subsection{Design Principles and Computational Considerations}\label{S3.1}

    The proposed NEFT architecture addresses the challenges of near-field CSI feedback by establishing a hierarchical framework that balances computational efficiency with reconstruction accuracy. As the foundational model in a family of Transformer-based networks, NEFT employs multi-stage downsampling and upsampling to achieve efficient feature compression while maintaining the flexibility to accommodate various device configurations and constraints. This design departs from conventional approaches by leveraging the ViT architecture's global attention capabilities to capture the complex spatial dependencies inherent in near-field scenarios.

    Near-field CSI feedback networks required to possess dual capabilities: capturing intricate local spatial details while modeling complex global dependencies across the channel matrix. ViT provides a promising solution through its global self-attention mechanism that can adaptively model long-range spatial relationships, overcoming the limitations of CNN-based approaches constrained by local receptive fields. The content-adaptive nature of Transformers, which is free from locality assumptions and translation equivariance constraints, makes them inherently suitable for learning the unique spatial structures characteristic of near-field scenarios.

    However, the quadratic computational complexity of standard self-attention mechanisms poses significant challenges for CSI feedback systems, particularly given the stringent real-time processing requirements imposed in wireless communications. Inspired by the hierarchical design principles of Swin Transformer\cite{swincfnet}, NEFT adopts a multi-stage architecture that strategically balances global attention capabilities with computational efficiency.
		
    Unlike W-MSA approaches, such as SwinCFNet, NEFT employs global attention across all hierarchical stages, implemented via global multi-head self-attention (MSA).
		This design is motivated by two key insights: (i) CSI matrices possess relatively small dimensions that remain computationally manageable after downsampling; (ii) the inherent global correlation structure of near-field CSI benefits more from spatially unrestricted attention than from locally constrained windows. Through progressive downsampling and upsampling across stages, NEFT substantially reduces computational overhead while directly capturing long-range spatial dependencies.

    \begin{figure}[!tbp]
		\vspace*{-2mm}
          \centering
          \captionsetup[subfigure]{font=footnotesize,labelfont=rm,textfont=rm}
          \subfloat[Patch merging]{%
            \includegraphics[width=0.5\columnwidth]{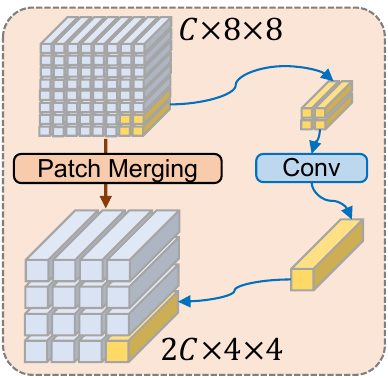}%
            \label{fig:patch-merge}%
          } 
           \hfill
          \subfloat[Patch division]{%
            \includegraphics[width=0.5\columnwidth]{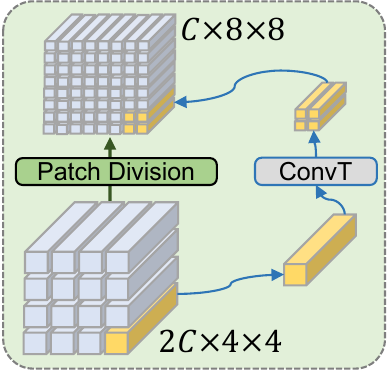}%
            \label{fig:patch-divide}%
          }
          \caption{Patch operations in NEFT. Merging aggregates 4 tokens to 1 via convolution (Conv); division reconstructs 1 to 4 tokens via transposed convolution (ConvT).}
          \label{fig:patch_op} 
					\vspace*{-3mm}
        \end{figure}
     
\subsection{Progressive Multi-Stage Architecture}\label{S3.2}

    As shown in \Cref{fig:NEFT}, the proposed NEFT framework employs a hierarchical Vision Transformer (HViT) architecture, adopting a two-stage encoder-decoder structure. 
    This hierarchical design reduces the number of tokens, lowering the quadratic complexity of self-attention ($O(N^2 d)$, where $N$ is the number of tokens and $d$ is the token embedding dimension), while enabling multi-scale feature learning to capture both fine-grained local variations and broader spatial dependencies in near-field CSI.

    The NEFT encoder partitions the input CSI tensor $(C,H,W) = (2,32,32)$ into $8 \times 8$ non-overlapping patches of size $2 \times 4 \times 4$, projecting each patch to an embedding dimension $C$ to form the input token array for the first ViT block. As illustrated in \Cref{fig:patch_op}(a), after the first ViT block, patch merging aggregates $2 \times 2$ neighboring tokens via a $2 \times 2$ convolution, reducing the number of spatial tokens and doubling the channel dimension to $2C \times 4 \times 4$. The coarse-grained tokens pass through a second ViT block and a linear projection to form the encoder's codeword. Due to hardware asymmetry, the encoder dimension is set smaller to reduce model complexity ($C = 32$).

    The decoder starts from the reshaped codeword ($2C \times 4 \times 4$) as input to the first ViT block. After this block, as illustrated in \Cref{fig:patch_op}(b), patch division via $2 \times 2$ transposed convolution (ConvT) inverts patch merging, expanding tokens to $C \times 8 \times 8$ for the second block. A final patch division reconstructs the original $2 \times 32 \times 32$ CSI matrix. The decoder dimension is larger ($C = 48$) to leverage more computational resources and improve reconstruction fidelity. For $\gamma = 64$, it is adjusted to $C = 40$ to maintain comparable parameter scale with baselines, while the proposed framework itself does not target specific parameter optimization.

\subsection{Computational Efficiency Analysis}\label{S3.3}

\begin{figure}[!b]
      \vspace*{-6mm}
			\begin{center}
      \includegraphics[width=0.45\textwidth]{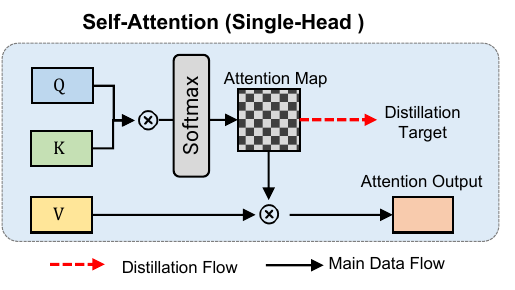}
			\end{center}
      \vspace*{-6mm}
      \caption{Self-attention computation process for a single attention head. The generated attention map is utilized for both feature weighting and knowledge distillation to the student network.}
      \label{fig:MSA} 
    \end{figure}

    NEFT employs MSA throughout all stages to maintain full receptive fields. 
		Fig.~\ref{fig:MSA} illustrates the computation flow of a single self-attention head.
		For an input feature map $\mathbf{X}\in\mathbb{R}^{C\times H\times W}$ flattened to the sequence representation $\mathbf{X}\in\mathbb{R}^{L_{t}\times C}$, where $L_{t}=H\times W$, the query, key, and value matrices are computed through linear projections
    \begin{equation}\label{eq8}
      \mathbf{Q}=\mathbf{X}\,\mathbf{W}^Q,\quad
      \mathbf{K}=\mathbf{X}\,\mathbf{W}^K,\quad
      \mathbf{V}=\mathbf{X}\,\mathbf{W}^V,
    \end{equation}
    where $\mathbf{W}^Q,\mathbf{W}^K,\mathbf{W}^V\in\mathbb{R}^{C\times C}$ are learnable parameters. The MSA, consisting of $h$ attention heads, is computed as
    \begin{equation}\label{eq9}
      \mathrm{MSA}(\mathbf{Q},\mathbf{K},\mathbf{V})
      =\mathrm{Concat}\bigl(\mathrm{head}_1,\dots,\mathrm{head}_h\bigr)\,\mathbf{W}^O,
    \end{equation}
    where each attention head is computed as $\mathrm{head}_i=\mathrm{softmax}\bigl(\mathbf{Q}_i\mathbf{K}_i^T/\sqrt{d_k}+\mathbf{B}\bigr)\,\mathbf{V}_i$, with $d_k=C/h$ denoting the dimension per head and $\mathbf{B}$ the learnable relative position bias. Here, $\mathbf{Q}_i,\mathbf{K}_i,\mathbf{V}_i$ are the projected queries, keys, and values for the $i$-th head. 
		As shown in Fig.~\ref{fig:MSA}, the attention maps from each head serve two roles: weighting the value matrices and providing supervision to the student network via the distillation mechanism described in the next section.

    Using larger patch embeddings ($4 \times 4$) , NEFT reduces token sequences from 256 and 64 tokens to 64 and 16 tokens, rendering MSA computationally feasible. This substantial reduction in sequence length is particularly beneficial given the quadratic complexity of attention mechanisms, while the information loss remains acceptable for CSI representation with typical token dimensions around 40.

    The computational advantages can be quantified through complexity analysis between W-MSA and MSA. The complexities are formalized as
    \begin{equation}\label{eq:win_comp} 
      \Omega(\text{W-MSA}) = 4hwC^2 + 2M^2 hw C,
    \end{equation}
    \begin{equation}\label{eq:glob_comp} 
      \Omega(\text{MSA}) = 4hwC^2 + 2 (hw)^2 C,
    \end{equation}
    where $h \times w$ denotes feature map dimensions, $C$ the channel dimension, and $M$ the window size. 
		
		W-MSA restricts attention to local windows and therefore requires multiple stacked blocks with shifted windows to approximate global context. For example, SwinCFNet uses $N_1=2$ and $N_2=4$ blocks per stage at $16\times16$ and $8\times 8$ resolutions, resulting in approximately 3.44\,M and 6.72\,M operations per stage.
		In contrast, with the same embedding dimension ($C = 40$), NEFT applies full MSA at coarser resolutions ($8\times 8$ and $4\times 4)$ with only $N_1 = N_2 = 1$, requiring 0.737\,M and 0.451\,M operations per stage. 
		Hence, a full MSA layer in NEFT consumes only 11.7\% of the computation of the W-MSA stacks in SwinCFNet, demonstrating that full attention is feasible once token counts are sufficiently reduced.
	
    \begin{figure*}[!t]
      \begin{center}
      \includegraphics[width=\textwidth]{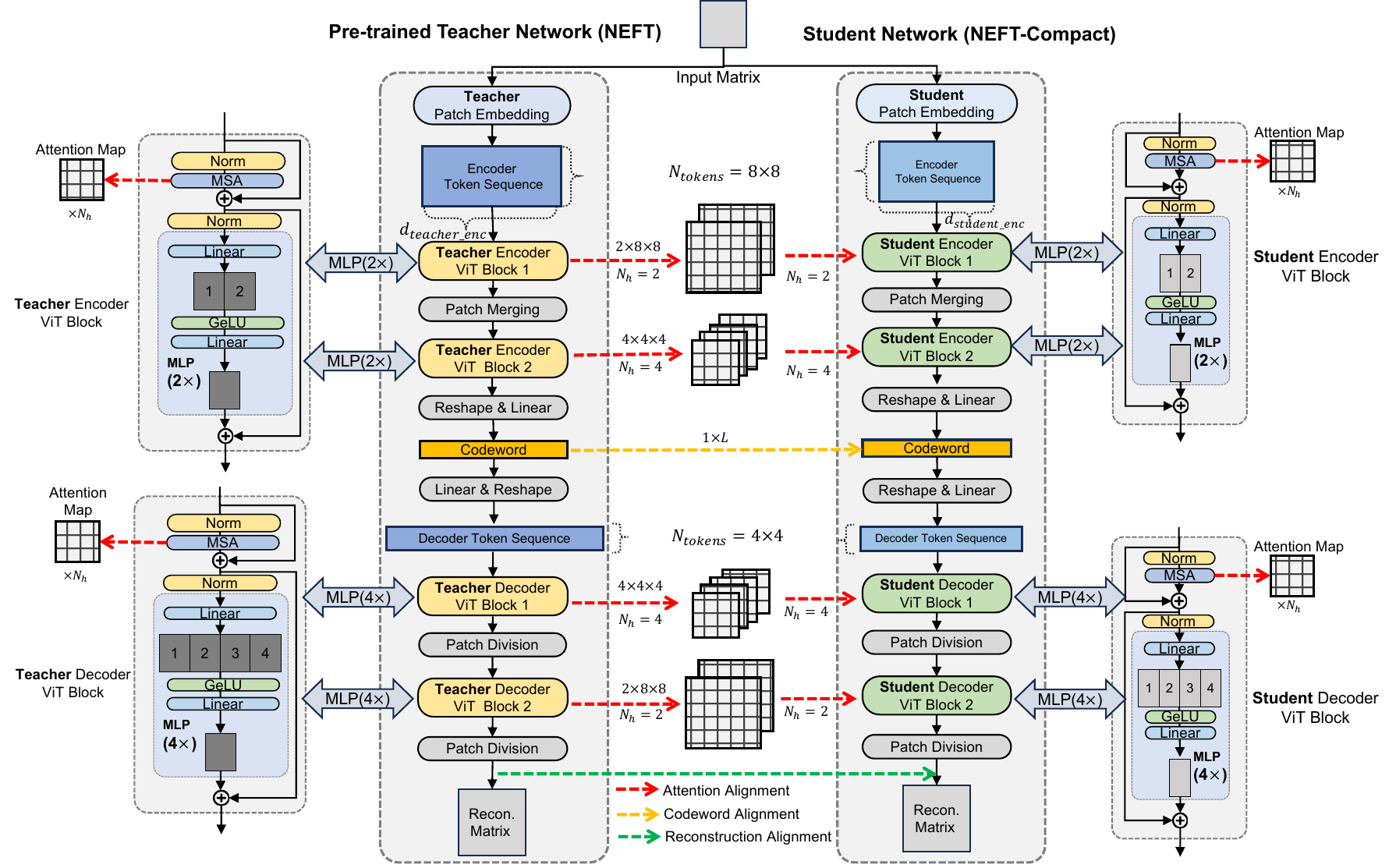}
			\end{center}
			\vspace*{-4mm}
      \caption{Overview of the proposed multi-level KD framework: the pre-trained teacher provides reconstruction outputs, attention maps, and bottleneck codewords from the encoder output as three complementary supervisory signals to guide the lightweight student network.}
      \label{fig:KD} 
			\vspace*{-4mm}
    \end{figure*}
    
\section{Multi-Level KD Framework for NEFT-Compact}\label{s4}

    This section introduces a multi-level KD framework designed to reduce the computational complexity of NEFT while retaining its reconstruction performance. The framework produces NEFT-Compact, a lightweight variant tailored for deployment under resource constraints.

\subsection{Framework Architecture and Design Rationale}\label{S4.1}

    The design of our multi-level KD framework is guided by three properties specific to the CSI feedback task:
    \begin{enumerate}
    \item CSI reconstruction is a regression task, where direct output alignment serves as a more suitable supervisory signal than the soft-label imitation used in classification.
    \item The spatial dependencies encoded by the MSA mechanism provide structured guidance for learning correlation-aware representations.
    \item Since the codeword directly encodes the compressed channel information, aligning this latent representation is crucial for downstream reconstruction accuracy.
    \end{enumerate}

    Accordingly, as illustrated in \Cref{fig:KD}, we extract three forms of supervision from the frozen teacher network: the final reconstruction output, the intermediate attention maps, and the bottleneck codeword. The final output provides the regression-specific ``dark knowledge,'' with the teacher's predictions serving as high-quality targets. The attention maps distill the teacher's internal representation of spatial correlations, enabling structural knowledge transfer. Lastly, the codeword provides a compact latent supervision signal, ensuring the student preserves essential information in the compressed domain. This multi-level strategy leverages supervision across the entire teacher network.

    A pre-trained NEFT model serves as the frozen teacher throughout the distillation. The student network, NEFT-Compact, is constructed as a width-reduced and structurally aligned counterpart to the teacher, enabling direct feature-level supervision. Both networks share an identical number of ViT blocks and attention heads, ensuring full architectural compatibility for layer-to-layer knowledge transfer. Model complexity is reduced exclusively by scaling down the token embedding dimension $d$, preserving depth and block topology.

    As shown in \Cref{fig:KD}, the teacher follows the asymmetric configuration described in Section~\ref{s3}, with its encoder operating at a lower embedding dimension than its decoder. To highlight width difference, the figure presents tokens in their flattened form $N_{tokens} \times d_{teacher}$, where $N_{tokens} = H \times W$, equivalent to the $C \times H \times W$ tensor view in ~\Cref{s3}. 
    The student network employs a reduced token width in both its encoder and decoder ($d_{student_{enc}} < d_{teacher_{enc}}$, $d_{student_{dec}} < d_{teacher_{dec}}$), resulting in thinner token sequences in all ViT blocks. This width reduction directly lowers the computational complexity, which is most significant in the MLP submodules due to their quadratic dependency on the token dimension.
    
    To instantiate this width reduction in practice, the decoder token dimension is uniformly reduced by 8 channels for all compression ratios $\gamma$. The encoder token dimension is reduced by $\{4, 8, 8\}$ channels under $\gamma=\{16, 32, 64\}$, respectively, since the baseline model at $\gamma=16$ operates at a relatively larger width. This rule is adopted to keep the student comparable in parameter scale to the teacher and baselines rather than to perform dimension-specific optimization.

    The student architecture is intentionally kept minimal to isolate the effect of the proposed KD mechanism rather than architectural re-design. Following established KD practices, the model complexity gap is introduced solely through width scaling while preserving depth and internal block structure\cite{hinton_kd, touvron_kd}. This width-only scaling setup ensures that performance gains arise from the multi-level KD rather than manual architectural tuning, which is further validated by the ablation results presented in Section~\ref{s6}.

\subsection{Multi-Level Alignment Mechanisms}\label{S4.2}

    The KD framework employs three complementary alignment strategies, each targeting a distinct representational level of the teacher network.

    \subsubsection{Reconstruction Alignment (RA)}
        RA aligns the final reconstruction outputs of the teacher and student networks to accelerate convergence and enhance stability. Instead of relying solely on the ground-truth CSI, the student benefits from the teacher’s refined reconstructions, which serve as achievable intermediate targets given the student’s limited capacity. This alignment bridges the performance gap by providing stable supervision throughout training.
        The RA loss is defined as
        \begin{equation}\label{eqRAloss} 
          \mathcal{L}_{RA} = \|\hat{\mathbf{H}}_{\text{NEFT}} - \hat{\mathbf{H}}_{\text{compact}}\|_2^2,
        \end{equation}
        where $\hat{\mathbf{H}}_{\text{NEFT}}$ and $\hat{\mathbf{H}}_{\text{compact}}$ denote the reconstructed CSI matrices from the teacher and student networks, respectively.

    \subsubsection{Attention Alignment (AA)}
        AA transfers the structured spatial correlations captured by the teacher’s MSA to the student. The teacher network, with its higher representational capacity, learns richer attention maps that encode more detailed inter-token dependencies. These patterns provide valuable supervision for improving the student’s spatial feature modeling.
    
        The framework adopts a layer-wise correspondence between teacher and student MSA modules, enabling direct alignment at each layer. For attention maps $\mathbf{A}_{\text{NEFT}}$ and $\mathbf{A}_{\text{compact}}$ with dimensions $[B, N_h, N_t, N_t]$, the alignment loss is
        \begin{equation}\label{eq:attention_map_alignment} 
          \mathcal{L}_{AA} = \frac{1}{{L}_{MSA}}\sum_{l=1}^{L_{MSA}}\frac{1}{N_h^{(l)}}\sum_{i=1}^{N_h^{(l)}} \|\mathbf{A}_{\text{NEFT}}^{(l,i)} - \mathbf{A}_{\text{compact}}^{(l,i)}\|_2^2,
        \end{equation}
        where ${L}_{MSA}$ is the number of MSA layers, $N_h^{(l)}$ is the number of heads in layer $l$, and $\mathbf{A}_{\text{NEFT}}^{(l,i)}, \mathbf{A}_{\text{compact}}^{(l,i)} \in \mathbb{R}^{N_t \times N_t}$ are the attention matrices for layer $l$ and head $i$, respectively. 
 
    \begin{figure*}[!t]
		\vspace*{-2mm}
      \begin{center}
      \includegraphics[width=\textwidth]{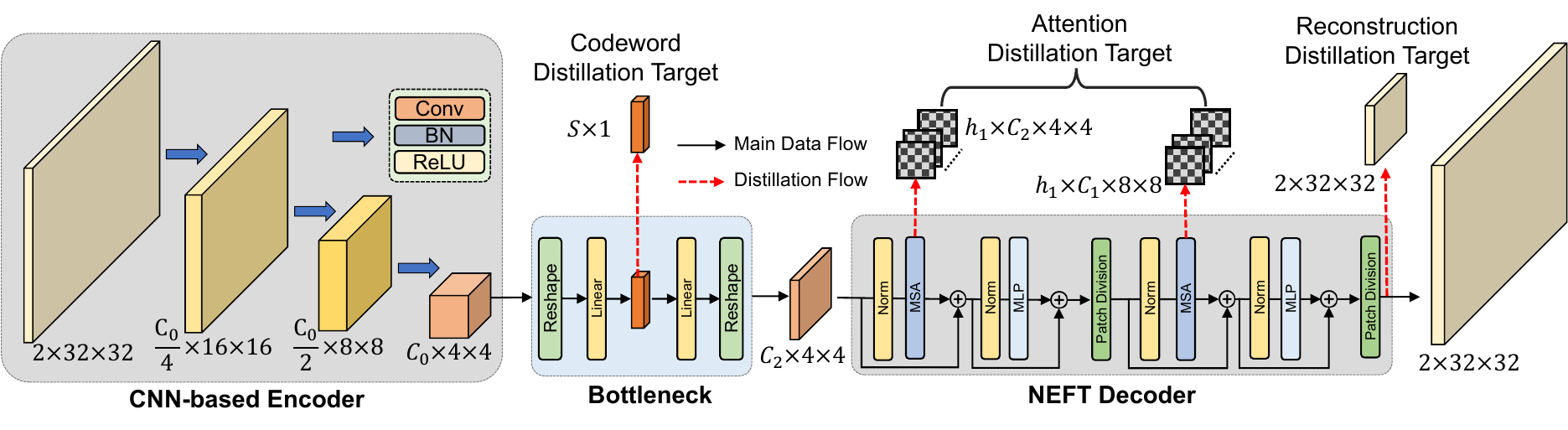}
			\end{center}
			\vspace*{-6mm}
      \caption{Architecture of NEFT-Hybrid combining lightweight CNN-based encoder with ViT-based decoder. The encoder employs progressive downsampling with channel expansion to extract local features efficiently, while the decoder leverages Transformer blocks to reconstruct global spatial correlations from compressed representations.}
      \label{Hybrid} 
			\vspace*{-4mm}
    \end{figure*}
			
    \subsubsection{Codeword Alignment (CA)}
        CA focuses on the encoder’s compressed codeword representations, which serve as the key information carriers for reconstruction. The teacher network, with greater capacity, learns more discriminative codewords that preserve essential channel characteristics. By constraining the student’s codewords to closely match those of the teacher, CA improves both the encoding process and the fidelity of the subsequent decoding.
				The CA loss is formulated as
        \begin{equation}\label{eq:codeword_alignment} 
        \mathcal{L}_{CA} = \frac{1}{N_d}\|\mathbf{z}_{\text{NEFT}} - \mathbf{z}_{\text{compact}}\|_2^2,
        \end{equation}
        where $\mathbf{z}_{\text{NEFT}}$ and $\mathbf{z}_{\text{compact}} \in \mathbb{R}^{N_d \times 1}$ are the codewords from the teacher and student encoders, respectively, and $N_d$ is the codeword dimension. This alignment stabilizes training and enhances the quality of the learned compressed representations.

\subsection{Training Strategy}\label{S4.3}

    The training strategy consists of two sequential phases: teacher pretraining and student distillation. The teacher network is first trained using conventional CSI feedback reconstruction objectives, minimizing the mean squared error (MSE) between original and reconstructed channel matrices until convergence. Once trained, the teacher parameters are frozen to provide stable supervision signals for the subsequent distillation phase.
    
    During student training, each input batch propagates through both the frozen teacher (inference mode) and trainable student networks. The student optimization combines reconstruction loss with three distillation objectives
    \begin{equation}\label{eq:total_loss} 
    \mathcal{L}_{\text{total}} = \mathcal{L}_{\text{rec}} + \lambda_1 \mathcal{L}_{RA} + \lambda_2 \mathcal{L}_{AA} + \lambda_3 \mathcal{L}_{CA},
    \end{equation}
		where the weighting coefficients are set as $\lambda_1 = 0.3$, $\lambda_2 = 2.0$ and $\lambda_3 = 2.0$ based on empirical analysis. The conservative weight for RA prevents optimization instability, while the larger weights for AA and CA compensate for their smaller numerical scales.
    
\section{NEFT-Hybrid: Hardware-Aware Architecture}\label{s5}

    This section introduces NEFT-Hybrid, which incorporates a lightweight CNN-based encoder to address the computational constraints of edge deployment, and derives NEFT-Edge through the proposed KD framework for ultra-efficient IoT applications.

\subsection{Motivation for Hybrid Architecture}\label{S5.1}
 
    The IoE paradigm in next-generation wireless systems requires CSI feedback frameworks that accommodate heterogeneous device capabilities. In practical deployments, the decoder operates at the BS side, where computation is less restricted, whereas the encoder runs on UE under strict latency and power limits. 
    NEFT-Compact improves efficiency through hierarchical design and multi-level KD, yet its encoder remains ViT-based and thus inherits the cost of self-attention and feed-forward projection. Self-attention increases computation due to pairwise token interaction, while the MLP layers further contribute to both computation and parameter storage, which affects real-time execution on edge hardware. While KD alleviates model size, it does not alter the encoder’s attention structure, and the token dimension cannot be substantially reduced without compromising representation capacity. 
    This suggests that compression alone is insufficient, and structural modification is required at the encoder side to reduce complexity at the source.
		
\begin{figure}[h]
	\vspace*{-6mm}
  \begin{center}
  \captionsetup[subfigure]{font=footnotesize,labelfont=rm,textfont=rm}
  \subfloat[\label{fig:enc_layer2}]{%
    \includegraphics[width=0.5\columnwidth]{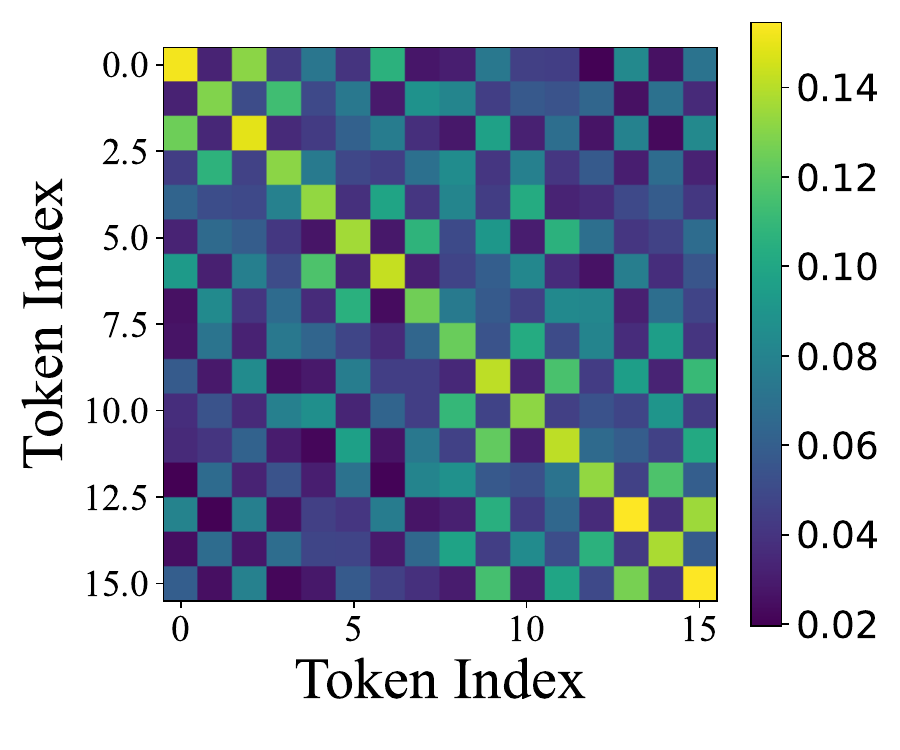}%
  }
  \hfill
  \subfloat[\label{fig:dec_layer1}]{%
    \includegraphics[width=0.5\columnwidth]{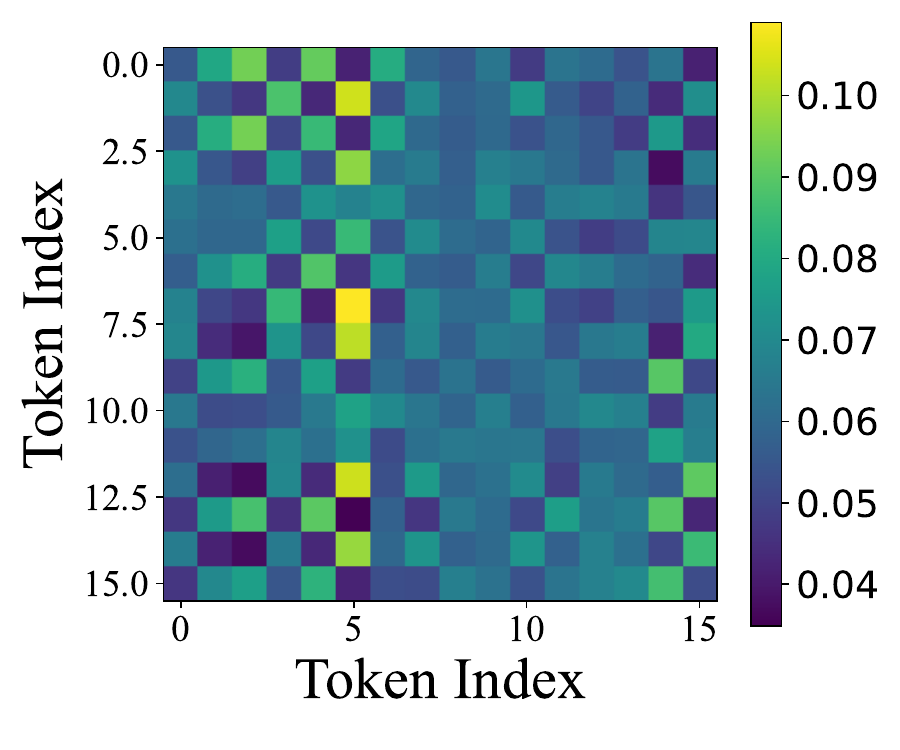}%
  }
	\end{center}
	\vspace*{-3mm}
  \caption{Attention map visualization of NEFT hierarchical stages.
    (a) Encoder stage-2 attention map;
    (b) Decoder stage-1 attention map.
    Coordinates represent patch indices, with intensity indicating attention strength.}
  \label{attn_map} 
	\vspace*{-2mm}
\end{figure}
 
      Attention visualization provides a concrete basis for this design choice. The attention maps are obtained by averaging the attention weights across all heads, providing a statistical view of token interaction patterns. 
      In an attention map, each element represents the correlation strength between two token positions determined by the row and column indices. Diagonal dominance indicates strong local dependencies, while a uniformly activated distribution reflects long-range interactions.
      As illustrated in \Cref{attn_map}\,(a), encoder stage-2 exhibits a strong diagonal attention pattern, indicating that feature interactions remain spatially localized. 
      Such locality suggests limited reliance on long-range token dependencies at the encoder side.
      In contrast, as shown in \Cref{attn_map}\,(b), decoder stage-1 shows a broadly distributed attention pattern with visible long-range activation, suggesting that global context modeling is primarily required in the decoding phase. 
 
    Prior work has shown that convolution can be formulated as a restricted form of local attention with fixed weights~\cite{conv_relationship, metaformer}. In this interpretation, a $3\times3$ convolutional kernel enforces a fixed receptive field around each spatial location, matching the diagonal locality revealed in the encoder attention map. Accordingly, we replace the ViT-based encoder in NEFT with a stack of three $3\times3$ convolutional layers, resulting in the proposed NEFT-Hybrid variant.

    The first convolutional layer replaces the patch embedding layer and generates an initial low-dimensional feature mapping. Unlike the patch embedding layer, which projects patches into a high-dimensional token space in a single step, the convolutional stack enables feature abstraction to be progressively established across layers, thereby reducing computation at the early encoder stage. 
    Subsequent convolution and spatial downsampling, combined with gradual channel expansion, enlarge the effective receptive field while keeping computation linear in spatial size. 
    In contrast to encoder-side processing, the ViT-based decoder is retained since long-range relational modeling remains necessary in the reconstruction stage, as indicated by the dispersed attention distribution. 
    This hybrid allocation of convolution for local encoding and attention for global decoding reduces encoder-side computation by more than 50\% while maintaining reconstruction fidelity, as validated in Section~\ref{s6}.

\subsection{NEFT-Hybrid and Ultra-Efficient Variant}\label{S5.2}

    In light of the localized and global attention patterns identified earlier, the following hybrid architecture aims to optimize both local feature extraction and long-range dependency modeling. 
    The proposed NEFT-Hybrid architecture, illustrated in \Cref{Hybrid}, employs a three-stage CNN encoder that extracts and compresses CSI features via hierarchical downsampling. 
    Each stage implements a progression where spatial dimensions are halved, the number of channels doubles, and representational depth increases to preserve information.
    The initial stage transforms the $32 \times 32$ input into $16 \times 16$ features with $C_0/4$ channels, where $C_0$ denotes the final encoding dimension. 
    Subsequent stages follow this pattern, producing a compact $4 \times 4 \times C_0$ representation for bandwidth-limited feedback.
    Here, we set $C_0 = \{60, 60, 56\}$ for $\gamma = \{16, 32, 64\}$, respectively. The reduction to 56 channels at $\gamma=64$ is used only to match the baseline model scale, not for optimization.

    Each encoding stage comprises a sequence of $3\times 3$ convolutional layers followed by batch normalization and ReLU activation. The $3\times 3$ kernels facilitate controlled receptive field expansion, which grows linearly with network depth while minimizing computational overhead. This design aligns with the localized correlation patterns observed in the encoder attention maps, enabling efficient capture of spatial redundancies through cascaded local operations. The computational complexity per layer is given by
    \begin{equation}\label{cnn_flops} 
      \mathrm{FLOPs} = H_{\text{out}} \times W_{\text{out}} \times C_{\text{out}} \times K_{h} \times K_{w} \times C_{\text{in}} ,
  \end{equation}
    where $H_{\text{out}}$ and $W_{\text{out}}$ denote the height and width of the output feature map, $C_{\text{out}}$ is the number of output channels, $K_{h}$ and $K_{w}$ represent the kernel height and width, respectively, and $C_{\text{in}}$ is the number of input channels. This configuration achieves approximately 50\% fewer FLOPs compared to the Transformer-based encoder in NEFT.
        
    The decoder maintains the HViT architecture from base NEFT, comprising hierarchical stages with multi-head self-attention and feed-forward networks. This configuration preserves the model's ability to reconstruct long-range spatial correlations from compressed representations—a critical requirement for accurate CSI recovery. By retaining ViT blocks in the decoder, NEFT-Hybrid leverages their superior capability in modeling global dependencies, as evidenced by the distributed attention patterns in \Cref{attn_map}(b). 
    The input token dimension for the decoder stages is denoted as $C_2$, set to $\{40, 40, 32\}$ for $\gamma = \{16, 32, 64\}$, with the reduction at $\gamma=64$ applied to match the baseline model scale.

    This architectural synergy enables practical deployment: the CNN encoder supports efficient on-device feature extraction, while the ViT-based decoder at the BS leverages available computational resources for high-fidelity reconstruction. 

    For ultra-constrained IoT and edge computing environments, however, even the reduced computational footprint of NEFT-Hybrid may remain prohibitive. To further adapt the model to such deployment scenarios, we apply the multi-level KD framework to compress NEFT-Hybrid and derive \textbf{NEFT-Edge}, a lightweight student network that maintains the hybrid encoder--decoder structure while operating at reduced token dimensionality. The distilled model transfers the teacher’s correlation-aware representations while reducing token dimensions in both encoder and decoder, achieving reconstruction accuracy close to the original NEFT-Hybrid with substantially lower computational cost than existing CNN-only near-field CSI feedback models.  

\section{Experimental Results} \label{s6}

    \subsection{Experimental Setup} \label{S6.1}

        To assess the performance of the proposed near-field CSI feedback framework, we adopt the LoS channel model \eqref{eq3} as in\cite{nf_channel_tcom}. The BS uses a ULA with $N_1\! =\! 1024$ antennas, and each UE has $N_2\! =\! 1$ antenna\cite{nf_cf_wcl}. We compute the Rayleigh distance $d_{\mathrm{R}}$ from \eqref{eq2}, then generate the dataset by uniform random sampling over $r\! \in\! [0.05\,d_{\mathrm{R}},\,0.5\,d_{\mathrm{R}}], \, \theta\! \in\! [0,\, 2\pi].$ Specifically, we produce 100,000 training samples, 10,000 validation samples, and 10,000 test samples to cover the entire near-field region. This LoS formulation is widely accepted as the standard reference model in near-field XL-MIMO studies\cite{tutorial_use_los,nf_channel_tcom}. 

        We benchmark our NEFT family against the two baseline methods, CsiNet\cite{csinet} and the state-of-the-art near-field CSI feedback model ExtendNLNet\cite{nf_cf_wcl}. The NEFT variants are:
        \begin{itemize}
          \item NEFT (Base Model)
          \item NEFT-Compact (Multi-level KD)
          \item NEFT-Hybrid (CNN-Transformer Hybrid Architecture)
          \item NEFT-Edge (Hybrid with Multi-level KD)
        \end{itemize}
  
    All networks are trained for 200 epochs using the AdamW optimizer. We initialize the learning rate at $1\times10^{-4}$ and employ a cosine-annealing schedule, which is defined as
    \begin{equation}\label{scheduler} 
      \eta_t = \eta_{\min} + \frac{1}{2}(\eta_{\max}-\eta_{\min})
        \bigl(1 + \cos\!\bigl(\tfrac{\pi t}{T}\bigr)\bigr),
    \end{equation}
    where $\eta_{\max}=1\times10^{-4}$, $\eta_{\min}=0$, and $T=200$ is the total number of epochs. Reconstruction quality is measured by the NMSE, which is defined as
    \begin{equation}\label{nmse} 
      \mathrm{NMSE(dB)} = 10\log_{10}\!\biggl(
          \frac{\mathbb{E}\bigl[\|\mathbf{H}-\hat{\mathbf{H}}\|_F^2\bigr]}
               {\mathbb{E}\bigl[\|\mathbf{H}\|_F^2\bigr]} \biggr),
    \end{equation}
    where $\|\mathbf{X}\|_F$ denotes the Frobenius norm of matrix $\mathbf{X}$. 
    We also evaluate the cosine similarity between the original and reconstructed CSI, defined as
    \begin{equation}\label{cosinesim} 
      \mathrm{\rho} = \mathbb{E}\!\left[
          \frac{\langle \mathbf{H}, \hat{\mathbf{H}}\rangle_F}
               {\|\mathbf{H}\|_F\,\|\hat{\mathbf{H}}\|_F} \right],
    \end{equation}
    where $\langle \mathbf{A}, \mathbf{B}\rangle_F = \mathrm{Tr}(\mathbf{A}^{\rm H} \mathbf{B})$ denotes the Frobenius inner product, and $(\cdot )^{\rm H}$ is the conjugate transpose operator.
  
    We apply early stopping so that training terminates if NMSE does not improve by at least 0.1\,dB over 20 consecutive epochs. For KD experiments, the initial learning rate is increased to $3\times10^{-4}$ to accommodate teacher-network guidance and the augmented loss function.
    
    All experiments are implemented in PyTorch 2.6.0 with CUDA 12.4.0 on an NVIDIA RTX 4080 GPU and an Intel Core i7-13700K CPU. Performance is reported in terms of NMSE and cosine similarity.

        \begin{table}[!b]
        \vspace*{-4mm}
        \caption{Performance Comparison Under Different Compression Ratios}
        \label{tab:performance_comparison} 
				\vspace*{-4mm}
        \setlength{\tabcolsep}{4pt}
        \renewcommand{\arraystretch}{1.1}
				\begin{center}
        \resizebox{\columnwidth}{!}{%
        \begin{tabular}{c l c c c c}
        \hline
        $\gamma$   & Model   & Parameters  & FLOPs (M)   & NMSE (dB)    & $\rho$ \\ \hline
        \multirow{4}{*}{16} 
            & ExtendNLNet \cite{nf_cf_wcl}  &543,456&13.66          & -13.19             & 97.92\%        \\
            & CsiNet \cite{csinet}          &\underline{530,656}&\underline{4.13}           & -10.06             & 94.86\%        \\
            & \textbf{Proposed NEFT}        &551,740     &6.35           &\textbf{-31.14}    & \textbf{99.94\%}        \\
            & \textbf{Proposed NEFT-Hybrid}  &\textbf{427,841}     &\textbf{4.07}  & \underline{-26.12}              & \underline{99.86\%}      \\ \hline
            \multirow{4}{*}{32} 
            & ExtendNLNet \cite{nf_cf_wcl}   &\underline{281,248}&13.40          & -11.19            & 96.68\%       \\
            & CsiNet \cite{csinet}           &\textbf{268,448}&\textbf{3.87}           & -8.40             & 92.40\%       \\
            & \textbf{Proposed NEFT}         &387,836     &6.18           & \textbf{-28.67}   & \textbf{99.91\%}       \\
            & \textbf{Proposed NEFT-Hybrid}  &284,417     &\underline{3.92}  & \underline{-24.29}           & \underline{99.80\%}       \\ \hline
            \multirow{4}{*}{64} 
            & ExtendNLNet \cite{nf_cf_wcl}  &\underline{150,144}&13.27           & -9.89              & 94.12\%       \\
            & CsiNet \cite{csinet}          &\textbf{137,344}&\underline{3.74}            & -6.77              & 88.73\%       \\
            & \textbf{Proposed NEFT}         &246,148     &4.82            & \textbf{-19.55}    & \textbf{99.45\%}       \\
            & \textbf{Proposed NEFT-Hybrid}  &181,596     &\textbf{3.22 }  & \underline{-17.82}             & \underline{99.25\% }      \\ \hline
        \end{tabular}%
        }
        \\[2pt]
        \footnotesize{CsiNet and ExtendNLNet use linear layers whose parameters shrink as $\gamma$ increases, weakening feature extraction. By contrast, our modules impose slightly more parameters at $\gamma\! =\! 64$ but achieving far better reconstruction.}
				\end{center}
				\vspace*{-3mm}
        \end{table}

    \subsection{Performance Evaluation}\label{S6.2}
		
        This subsection evaluates the performance of our NEFT and NEFT-Hybrid architectures against the two baseline methods across various compression ratios. Table~\ref{tab:performance_comparison} compares the four models, ExtendNLNet, CsiNet, NEFT and NEFT-Hybrid, under compression ratios $\gamma = \{16, 32, 64\}$, in terms of the NMSE \eqref{nmse}, the cosine similarity $\rho$ \eqref{cosinesim}, and computational cost. In Table~\ref{tab:performance_comparison}, the boldface number indicates the best value, and the underlined one indicates the second best.
						
        \begin{table*}[b!]
				\vspace*{-4mm}
				\begin{center}
        \renewcommand{\arraystretch}{1.2}
        \setlength{\tabcolsep}{4pt}
        \caption{Comprehensive Evaluation of the Proposed Distillation Framework on NEFT and NEFT-Hybrid}
        \begin{threeparttable}
        \label{tab:distillation_comprehensive} 
				\vspace*{-5mm}
                \begin{tabular}{c l | c c | c c c c | c c}
                \hline
                \multirow{2}{*}{$\gamma$} & \multirow{2}{*}{Model Configuration} 
                & \multicolumn{2}{c|}{Parameters} 
                & \multicolumn{4}{c|}{FLOPs (M)} 
                & \multicolumn{2}{c}{NMSE (dB)} \\
                \cline{3-10}
                & & Value & Red. (\%)
                & Total & Enc. & Total Red. (\%) & Enc. Red. (\%)
                & Value & Degr. (\%) \\
                \hline            
                    \multirow{8}{*}{16} 
                     & \textbf{Proposed NEFT}& 551,740 & --- & 6.35 & 1.72 & --- & --- & -31.14 & --- \\
                     & \textbf{Proposed NEFT-Compact}& 439,220 & \textbf{20.39} & 4.70 & 1.38 & \textbf{25.98} & \textbf{19.98} & \textbf{-30.13} &\textbf{ 3.22} \\
                     & NEFT-Compact-onlyRecon.& --- & --- & --- & --- & --- & --- & -28.98& 6.93\\
                     & NEFT-Compact-w/o-KD& --- & --- & --- & --- & --- & --- & -28.33& 9.00 \\
                    \cline{2-10}
                     & \textbf{Proposed NEFT-Hybrid} & 427,841 & --- & 4.07 & 0.74 & --- & --- & -26.12 & --- \\
                     & \textbf{Proposed NEFT-Edge}& 300,340 & \textbf{29.80} & 2.61 & 0.38 & \textbf{35.87} & \textbf{48.79} & \textbf{-24.24 }& \textbf{7.20 }\\
                     & NEFT-Edge-onlyRecon.& --- & --- & --- & --- & --- & --- & -23.18 & 11.26 \\
                     & NEFT-Edge-w/o-KD& --- & --- & --- & --- & --- & --- & -22.55 & 13.68 \\
                    \hline
                    \multirow{8}{*}{32}
                     & \textbf{Proposed NEFT}& 387,836 & --- & 6.18 & 1.66 & --- & --- & -28.67 & --- \\
                     & \textbf{Proposed NEFT-Compact}& 281,412 & \textbf{27.44} & 4.26 & 1.02 & \textbf{31.07} & \textbf{38.38} &\textbf{ -27.14}& \textbf{5.34}\\
                     & NEFT-Compact-onlyRecon.& --- & --- & --- & --- & --- & --- & -26.05& 9.15\\
                     & NEFT-Compact-w/o-KD& --- & --- & --- & --- & --- & --- & -25.09& 12.49\\
                    \cline{2-10}
                     & \textbf{Proposed NEFT-Hybrid}& 284,417 & --- & 3.92 & 0.68 & --- & --- & -24.29 & --- \\
                     & \textbf{Proposed NEFT-Edge}& 193,780 & \textbf{31.87} & 2.51 & 0.34 & \textbf{35.97 }& \textbf{50.15} & \textbf{-22.00} & \textbf{9.45} \\
                     & NEFT-Edge-onlyRecon. & --- & --- & --- & --- & --- & --- & -20.91 & 13.94 \\
                     & NEFT-Edge-w/o-KD & --- & --- & --- & --- & --- & --- & -20.30 & 16.43 \\
                    \hline
                    \multirow{8}{*}{64}
                     & \textbf{Proposed NEFT}& 246,148 & --- & 4.82 & 1.62 & --- & --- & -19.55 & --- \\
                     & \textbf{Proposed NEFT-Compact}& 162,408 & \textbf{34.02} & 3.06 & 0.92 & \textbf{36.51} & \textbf{43.10} & \textbf{-17.99}& \textbf{7.98}\\
                     & NEFT-Compact-onlyRecon.& --- & --- & --- & --- & --- & --- & -15.26& 21.94 \\
                     & NEFT-Compact-w/o-KD& --- & --- & --- & --- & --- & --- & -14.35& 26.59 \\
                    \cline{2-10}
                     & \textbf{Proposed NEFT-Hybrid} & 181,596 & --- & 3.22 & 0.58 & --- & --- & -17.82& --- \\
                     & \textbf{Proposed NEFT-Edge}& 140,500 & \textbf{22.63} & 2.45 & 0.32 & \textbf{23.91} & \textbf{44.52} & \textbf{-15.65}& \textbf{12.18}\\
                     & NEFT-Edge-onlyRecon.& --- & --- & --- & --- & --- & --- & -12.03& 22.49 \\
                     & NEFT-Edge-w/o-KD& --- & --- & --- & --- & --- & --- & -10.75& 39.66 \\
                    \hline
                \end{tabular}
        \begin{tablenotes}
        \footnotesize
            \item[a] ``Red.'' denotes the reduction ratio, ``Degr.'' denotes NMSE degradation relative to the teacher model, and ``Enc.'' refers to the encoder part of the model.
            \item[b] ``onlyRecon.'' denotes models trained with distillation using only the reconstruction loss (no multi-level alignment). ``w/o-KD'' denotes models trained without knowledge distillation.
            \item[c] ``---'' indicates a zero value or the same as the corresponding teacher model.
        \end{tablenotes}
        \end{threeparttable}
        \end{center}
				\vspace*{-1mm}
        \end{table*}
       
        For $\gamma = 16$, our NEFT achieves the best performance, significantly outperforming both ExtendNLNet and CsiNet in terms of NMSE and cosine similarity. In addition to this significant performance improvement, NEFT maintains a comparable parameter count as the previous state-of-the-art ExtendNLNet and requires less than half the computational cost of ExtendNLNet. NEFT-Hybrid, optimized for mobile deployment, also significantly outperforms ExtendNLNet and CsiNete in terms of NMSE and cosine similarity, and its performance is very close to that of NEFT. Furthermore, its computational cost and parameter count are the lowest among all the models. This demonstrates that NEFT-Hybrid achieves an excellent balance between performance and efficiency, making it well-suited for resource-constrained scenarios. 

        At higher compression ratios ($\gamma = 32$ and $\gamma = 64$), the proposed NEFT also achieves the best reconstruction performance, with NMSE substantially lower than both ExtendNLNet and CsiNet and with better cosine similarity than the two baselines. Also NEFT requires less than half the FLOPs needed by ExtendNLNet. Although NEFT has more parameters than the two baselines, it delivers considerably higher reconstruction quality, demonstrating a favorable balance among accuracy, computational cost and model size.
        
        NEFT-Hybrid remains highly efficient across both higher compression ratios, delivering competitive reconstruction quality to NEFT, while reducing computational cost to the level of CsiNet.
				The two baselines rely on linear layers for both feature extraction and dimensionality reduction. At high compression ratios, these layers shrink sharply in complexity and parameters but suffer from a substantial drop in feature extraction capability.
        In contrast, NEFT-Hybrid retains lightweight convolutional extractors, not only modestly reducing the number of parameters and the computational cost but also preserving reconstruction quality under extreme compression.
				
        The results demonstrate the performance advantages of NEFT and NEFT-Hybrid. NEFT consistently achieves the best reconstruction quality across all compression ratios while requiring far fewer FLOPs than ExtendNLNet, highlighting the capability of ViT for near-field CSI feedback and the effectiveness of the proposed hierarchical design. NEFT-Hybrid further balances accuracy and efficiency by pairing lightweight CNN encoders with the NEFT decoder, making it particularly well suited for mobile deployment.

    \subsection{Multi-Level KD Effectiveness} \label{S6.3}

        This subsection evaluates the effectiveness of the proposed multi-level KD framework on NEFT and NEFT-Hybrid architectures through comprehensive experiments. Table \ref{tab:distillation_comprehensive} summarizes the results, including comparisons of teacher models, distilled student models, and ablation studies with reconstruction-only and without-KD settings.
       
\begin{figure*}[!b]
\vspace*{-6mm}
\begin{center}
\subfloat[\label{component_loss}]{%
\includegraphics[width=0.9\columnwidth]{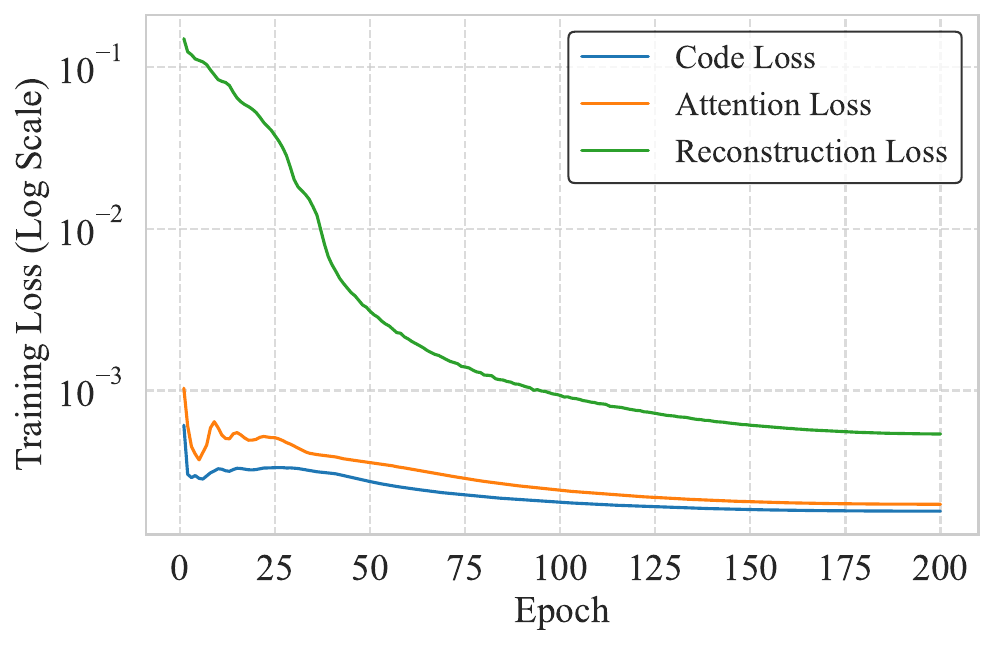}%
}
\hfill
\subfloat[\label{KD_Loss}]{%
\includegraphics[width=0.9\columnwidth]{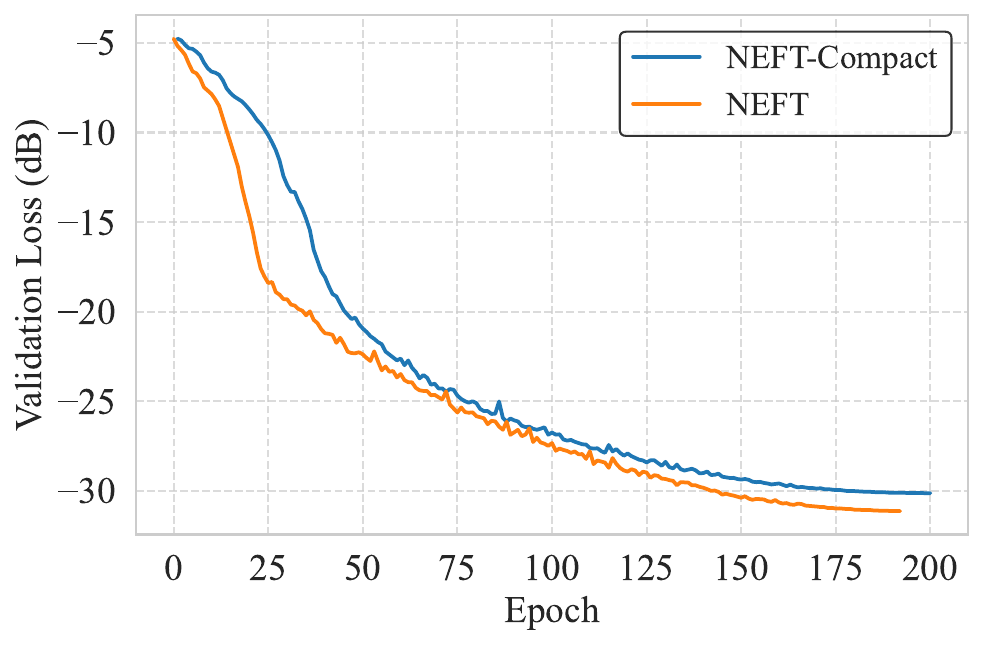}%
}
\end{center}
\vspace*{-2mm}
\caption{The loss curves during KD training. (a) Loss curves of individual components in the overall distillation loss, and (b) the validation loss curves for both teacher and student models.}
\label{fig:loss_analysis} 
\vspace*{-1mm}
\end{figure*}

        At $\gamma = 16$, the proposed multi-level KD framework enables substantial model compression while maintaining strong reconstruction performance. NEFT-Compact reduces parameters, total FLOPs and encoder FLOPs by 20.39\%, 25.98\% and 19.98\%, respectively, with only a 3.22\% NMSE degradation. NEFT-Edge achieves even larger reductions of 29.80\%, 35.87\% and 48.79\% in parameters, total FLOPs and encoder FLOPs, respectively, but accompanied by a 7.20\% NMSE degradation. Ablation results confirm the significance of KD: models trained without KD show noticeably higher NMSE degradations of 9.00\% for NEFT-Compact-w/o-KD and 13.68\% for NEFT-Edge-w/o-KD. The results also show that reconstruction-only training leads to inferior performance with NMSE degradations of 6.93\% and 11.26\% for NEFT-Compact-onlyRecon. and NEFT-Edge-onlyRecon., respectively.
				
        At $\gamma = 32$, the framework continues to provide favorable compression–accuracy tradeoffs. NEFT-Compact lowers parameters by 27.44\%, total FLOPs by 31.07\% and encoder FLOPs by 38.38\%, with a controlled NMSE degradation of 5.34\%. NEFT-Edge further reaches reductions of 31.87\%, 35.97\% and 50.15\% in parameters,  total FLOPs and encoder FLOPs, respectively, with a 9.45\% NMSE degradation. In contrast, the two models without KD experience substantially larger NMSE degradations of 12.49\% and 16.43\%, while the two models with reconstruction-only training also leads to considerable NMSE degradations of 9.15\% and 13.94\%.
				
				At the highest compression level $\gamma = 62$, the benefit of multi-level KD becomes most evident. NEFT-Compact achieves 34.02\%, 36.51\% and 43.10\% reductions in parameters, total FLOPs and encoder FLOPs, respectively, while keeping the NMSE degradation at 7.98\%. NEFT-Edge yields reductions of 22.63\%, 23.91\% and 44.52\% in parameters, total FLOPs and encoder FLOPs, with a 12.18\% NMSE degradation. The gap from the ablation models is particularly pronounced at this setting: NEFT-Compact-w/o-KD and NEFT-Edge-w/o-KD suffer severe NMSE degradations of 26.59\% and 39.66\%, and the two models with reconstruction-only training also results in substantial NMSE degradations of 21.94\% and 22.49\%.
				
        These results demonstrate that the proposed multilevel KD framework enables compact models to retain the majority of the teacher model's accuracy even under aggressive compression. The results also show that reconstruction-only training is insufficient, highlighting the crucial role of multi-level alignment in effectively transferring knowledge and stabilizing student performance across a wide range of compression ratios.
 
        \Cref{fig:loss_analysis} illustrates the training dynamics of our multi-level distillation approach. As shown in \Cref{fig:loss_analysis}\,(a), the individual components of the overall distillation loss converge stably during training. In \Cref{fig:loss_analysis}\,(b), the validation loss curves reveal that the student model initially converges more slowly than the teacher model, as the KD loss has limited influence in the early stages. With training progression, the KD loss effectively guides the student model, allowing its loss to closely approximate that of the teacher model. However, due to differences in network complexity, the student model exhibits a slightly higher loss in later stages. Nevertheless, the overall loss gap remains minimal, demonstrating the effectiveness of the distillation process in achieving competitive performance with a compact model.
 
       The loss curves in Fig.~\ref{fig:loss_analysis} further complement the performance analysis by showing that the student model maintains stable convergence throughout training. Although the student initially converges more slowly, the multi-level distillation progressively aligns its intermediate representations with those of the teacher, resulting in a final loss that is close to the teacher's. These stable training dynamics, combined with the NEFT-Hybrid architecture's substantial encoder-side computation reduction, highlight the framework's suitability for mobile and other resource-limited deployments. Overall, the results demonstrate
that multi-level distillation delivers reliable training behavior together with meaningful efficiency gains, providing a strong solution for high-efficiency CSI feedback.
 
    \subsection{Computation–Accuracy Trade-off Analysis} \label{S6.4}
		
\begin{figure}[!tp]
				\vspace*{-1mm}
        \begin{center}
        \includegraphics[width=0.42\textwidth]{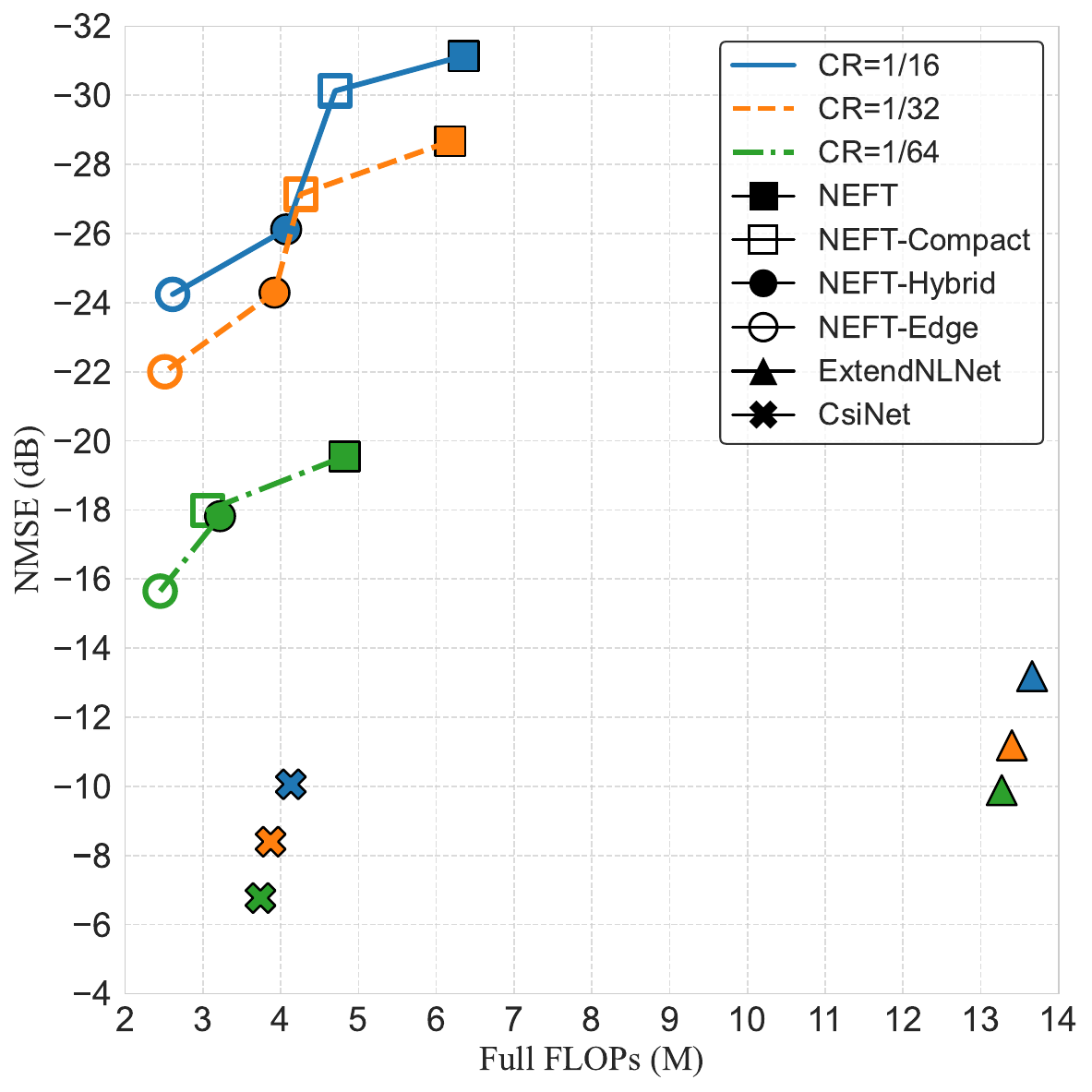}
				\end{center}
				\vspace*{-6mm}
        \caption{Trade-off between reconstruction performance and total model complexity (FLOPs) across different architectures.}
        \label{fig:efficiency_full} 
				\vspace*{-2mm}
        \end{figure}

\begin{figure}[!tp]
				\vspace*{-1mm}
        \begin{center}
        \includegraphics[width=0.42\textwidth]{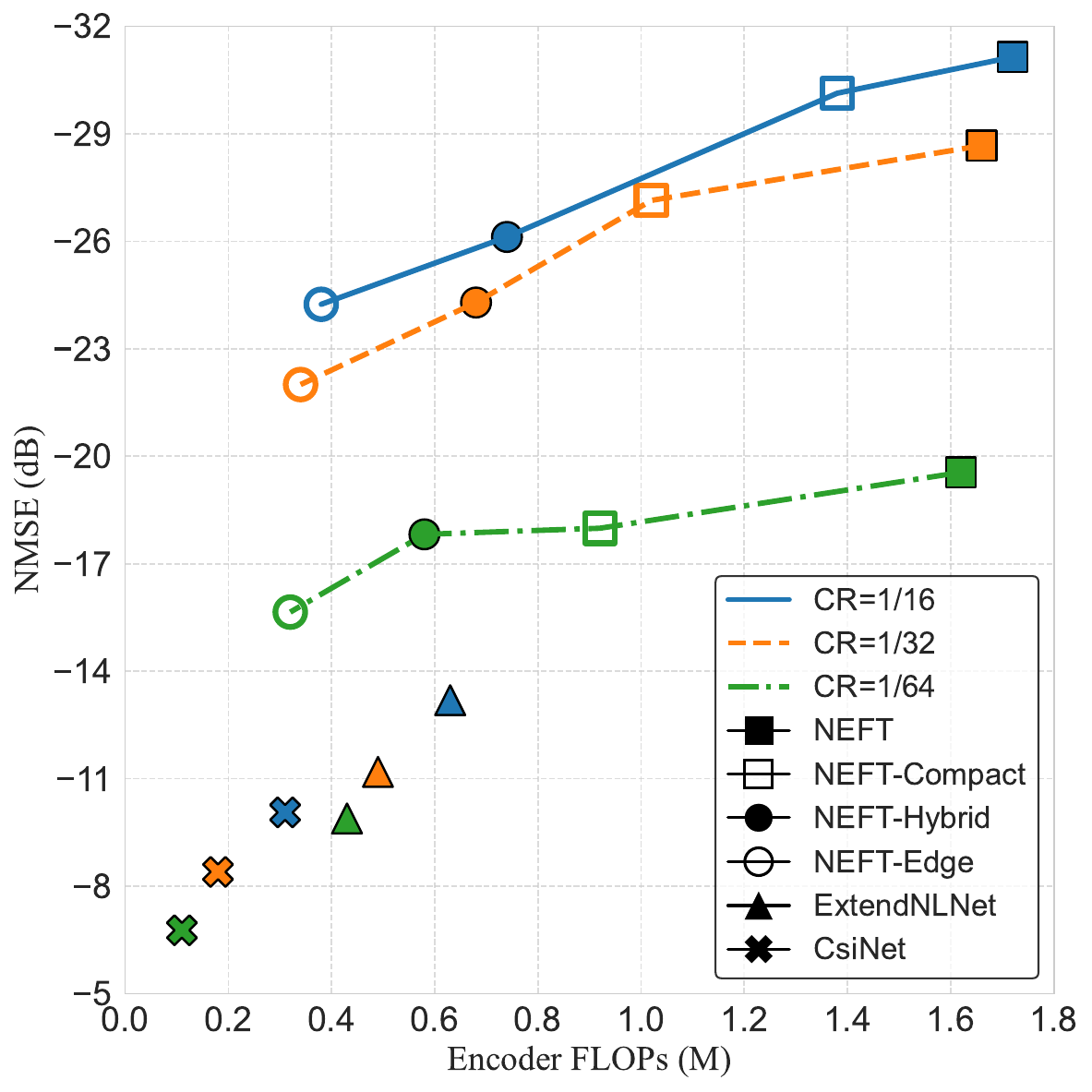}
				\end{center}
				\vspace*{-6mm}
        \caption{Trade-off between reconstruction performance and encoder-side computational complexity (FLOPs) across different models.}
        \label{fig:efficiency_encoder} 
				\vspace*{-4mm}
        \end{figure}

        To rigorously assess the trade-off between reconstruction performance and computational efficiency, we examine the relationship between model complexity and performance across multiple architectures and compression ratios. \Cref{fig:efficiency_full} illustrates the overall model-level trade-offs between reconstruction performance and computational complexity, while \Cref{fig:efficiency_encoder} presents encoder-side complexity versus reconstruction quality. In the both plots, points closer to the upper-left corner indicate superior efficiency–performance trade-offs, i.e., lower computational cost and higher reconstruction accuracy. The evaluation includes two baseline models (CsiNet and ExtendNLNet) as well as the proposed NEFT models at three compression ratios ($\gamma = 16, 32, 64$).
				
        As shown in \Cref{fig:efficiency_full}, the total complexity figure reveals distinct efficiency frontiers for different deployment scenarios. The NEFT family defines clear frontiers in terms of overall model efficiency. At $\gamma = 16$, the NEFT base model achieves the highest reconstruction quality with a balanced computational footprint, significantly outperforming ExtendNLNet in both accuracy and resource utilization. The combination of multi-level KD and the CNN-based hybrid encoder enables the lightweight variants to occupy the mid-efficiency region. From high-performance models to NEFT-Edge, all the proposed architectures establish new state-of-the-art efficiency–performance boundary.
        
        As the compression ratio increases, the advantages of our designs remain evident, although the performance gap between NEFT-Compact and NEFT-Hybrid narrows. This is mainly due to tighter computational constraints at high compression levels, which require smaller input patch sizes for Transformer encoders. However, smaller patches reduce the available feature context for self-attention, limiting the ability to capture long-range dependencies. In contrast, CNN encoders, which specialize in local feature extraction, maintain more consistent performance under these conditions. Consequently, the performance of Transformer-based encoders approaches that of CNN-based models, underscoring the robustness and adaptability of the hybrid design under stringent constraints.
        
        In mobile deployment scenarios, encoder-side complexity becomes a particularly critical factor. As illustrated in \Cref{fig:efficiency_encoder}, although the NEFT base model exhibits higher encoder complexity than the two baselines, its nearly 20\,dB reconstruction accuracy gain, together with the increasing computational capabilities of modern devices, make it well suited for high-performance mobile platforms. Moreover, the multi-level KD framework and hybrid architecture deliver significant efficiency gains. NEFT-Edge, in particular, achieves both lower encoder complexity and better performance than ExtendNLNet, demonstrating the effectiveness of the proposed lightweight strategy.
  
\section{Conclusions}\label{S7}

    This paper proposes NEFT, a unified framework for near-field CSI feedback in XL-MIMO systems, which integrates Transformer-based architectures, model compression and hybrid design strategies. NEFT demonstrates remarkable reconstruction quality, achieving a 15-21\,dB improvement in reconstruction NMSE compared to existing state-of-the-art methods, while simultaneously reducing computational complexity. A multi-level KD framework enables deployment across heterogeneous platforms, with NEFT-Compact and NEFT-Edge reducing total FLOPs by 25–36\% without notable accuracy loss. In addition, the hybrid encoder–decoder architecture further reduces encoder-side complexity by up to 64\%, making NEFT-Hybrid suitable for asymmetric device scenarios. These results establish NEFT as a practical and scalable solution for efficient CSI feedback in near-field conditions.

\bibliographystyle{IEEEtran}  


\end{document}